\documentclass[twocolumn,amsmath,amssymb]{revtex4}

\usepackage{epstopdf}
\usepackage{graphicx}
\usepackage{dcolumn}
\usepackage{bm}
\usepackage{color}

\begin{document}
\title{Prospects for a direct dark matter search using high resistivity CCD detectors}

\author{J. Estrada$^1$,  H. Cease$^{1}$, H.T. Diehl$^{1}$, B. Flaugher$^{1}$, J. Jones$^{1,2}$, D. Kubik$^1$ and A. Sonnenschein$^1$}
\address{ $^{1}$Fermi National Accelerator Laboratory, Batavia, Illinois 60510, USA \\
                  $^{2}$Batavia High School, Batavia, Illinois 60510, USA     \\}

\date{\today}

\begin{abstract}
The possibility of using CCD detectors in a low threshold direct detection dark matter search experiment is discussed. We present
the main features of the DECam detectors that make them a good alternative for such an experiment, namely their
low noise and their large depleted volume. The performance of the DECam CCDs for  the detection 
of nuclear recoils is discussed, and a measurement of the ionization efficiency for these events is presented. 
Finally the plans and expected reach for the CCD Experiment at Low Background (CELB) are discussed.
\end{abstract}

\maketitle

\section{Introduction}

The current results from observations in astronomy and
astrophysics strongly favor the concordance model for cosmology \cite{concordance}, where
$\sim$25$\%$ of the total energy in the universe is in the form of
dark matter. Direct detection of dark matter is a possible
interpretation for the annual modulation signal in the 
DAMA/LIBRA   and DAMA/NaI 
experiments  described in \cite{damalibra} and references therein, but has not yet been confirmed by other. The
only confirmed evidence for dark matter  comes from its gravitational
effects in space.  The imminent turn-on of the Large Hadron Collider (LHC) 
high energy physics collider could change this situation dramatically by confirming
the existence of new a particle having the properties needed
to be a good dark matter candidate. Such a discovery, however, will
not tell us if the new particle is indeed the dark matter unless 
we directly detect it in the Galactic halo. 

\begin{figure}
\begin{center}
\includegraphics[width=1.\columnwidth]{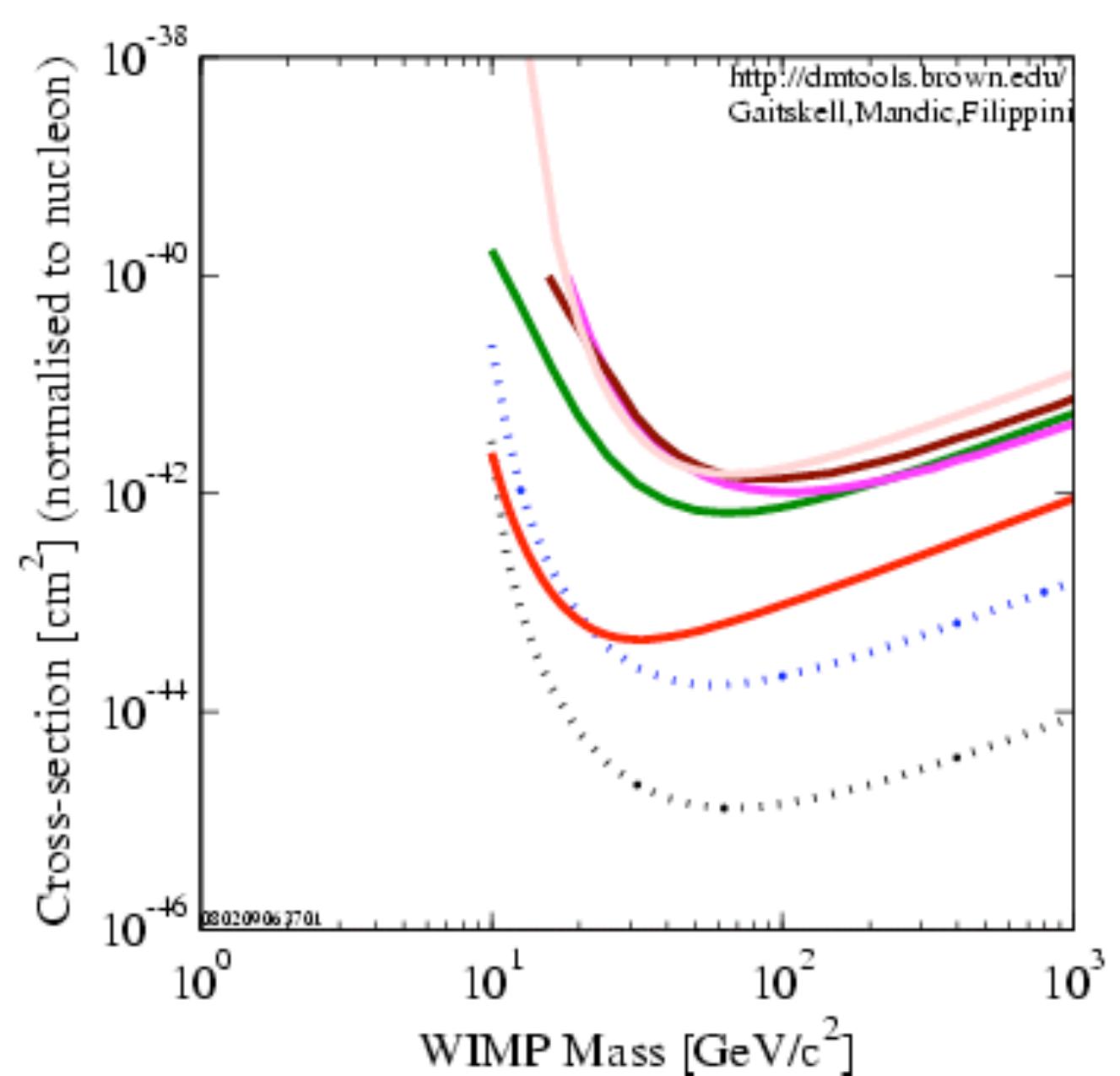}
\includegraphics[width=1.\columnwidth]{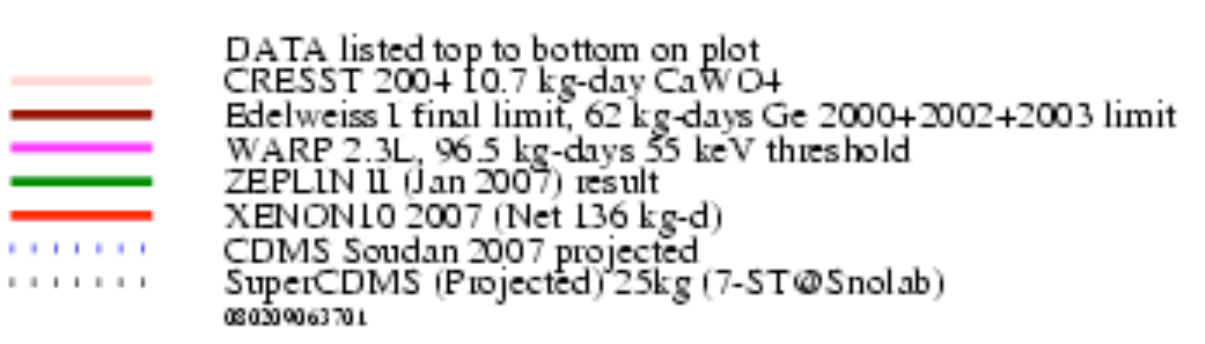}
\end{center}
\caption{Examples of cross section limits from existing and planned direct dark matter searches. The
limits are very weak for WIMP mass below 10 GeV. }
\label{fig:DM_limits}
\end{figure}

The Weakly Interactive Massive Particles (WIMP) are the leading candidate
dark matter particle. The search for dark matter in the halo of our galaxy by direct detection experiments 
has been a very active field in recent years, with the limits
on WIMP-nucleon cross section improving significantly with time (for a review see Refs. \cite{ddm1} and \cite{ddm2}). 
In general, these experiments are performed by measuring the rate of nuclear 
recoils in a detector above a certain energy threshold.
 By doing the experiments with appropriate shielding and in an
underground facility, a very low rate  ( $\approx$1 eV/kg/day ) 
for  nuclear recoils coming from non-WIMP events can be achieved.
These low measured rates are used to establish upper limits 
for the WIMP-nucleon cross section as a function of WIMP mass.
However, typical searches of this kind
have a poor sensitivity for low mass WIMPs because 
of the high thresholds ($\approx$1 keV) established for the detection of the nuclear
recoils.  These high thresholds produce  weaker experimental limits for WIMP masses
below 10 GeV, as can be seen in Fig. \ref{fig:DM_limits}.  
The most popular models for dark matter particles predict masses above 
10 GeV (for a review see Ref. \cite{susy_DM}) for which a low threshold
dark matter search is not needed. There are however
some models for which the dark matter particles have
lower mass \cite{lightDM0} \cite{lightDM1} and in those cases a low threshold
for nuclear recoil detection is needed for a direct search.  A low mass
dark matter candidate could also arise in the
most simple extension of the Minimal Supersymmetric Standard Model  
as discussed in Ref. \cite{lightDM5} .
The possibility of detecting sub-GeV dark matter candidates 
in colliders is discussed in  Ref. \cite{lightDM4}.
Models for which the typical velocity  of the dark matter particles
with respect to the earth is lower have also been considered \cite{lightDM2}  \cite{lightDM3},
and those cases also require a low threshold in a direct search experiment. 
One of the limitations for setting a lower
threshold in the direct dark matter searches is the readout noise of the detectors used
for the experiments. There has been a continuing effort in reducing the 
readout noise for the detectors typically used in direct dark matter searches 
to overcome the high threshold issue \cite{juancollar}.

In this work, we discuss the idea of using Charge-Coupled Devices (CCDs) 
for a direct dark matter search with
a low detection threshold for nuclear recoils. The CCDs considered here present an extremely 
low readout noise   $\sigma  = 2$e and could 
easily be stacked in an array to produce a detector with significant active mass.

\section{High resistivity CCD detectors}

Recent advances in CCD technology \cite{LBNL1}  allow the fabrication of
high resistivity ($\sim 10 k\Omega /$cm$^2$) detectors,   up to $\sim 300$ $\mu$m thick 
which are fully depleted at relatively low voltages. These CCDs have a 
significantly higher efficiency in the near-IR and for this reason
are the optical detectors chosen by
several groups building new mosaic cameras for astronomy, such as DECam \cite{DECam, DES} , 
SNAP \cite{SNAP} and  HyperSuprime \cite{Hypersuprime}. 

In this work we discuss the features of the 
DECam  detectors \cite{DECam, DECam_CCDs, DECam_CCDtest}
that will make then good candidates for a low threshold direct
dark matter search. DECam is the instrument currently being built for the Blanco 4m Telescope at CTIO 
\cite{BlancoCTIO} that will be used for the Dark Energy Survey (DES) and will be available
as a facility instrument at CTIO.
For a general description of CCDs see Ref \cite{janesick}.

\begin{figure}
\begin{center}
\includegraphics[width=0.7\columnwidth]{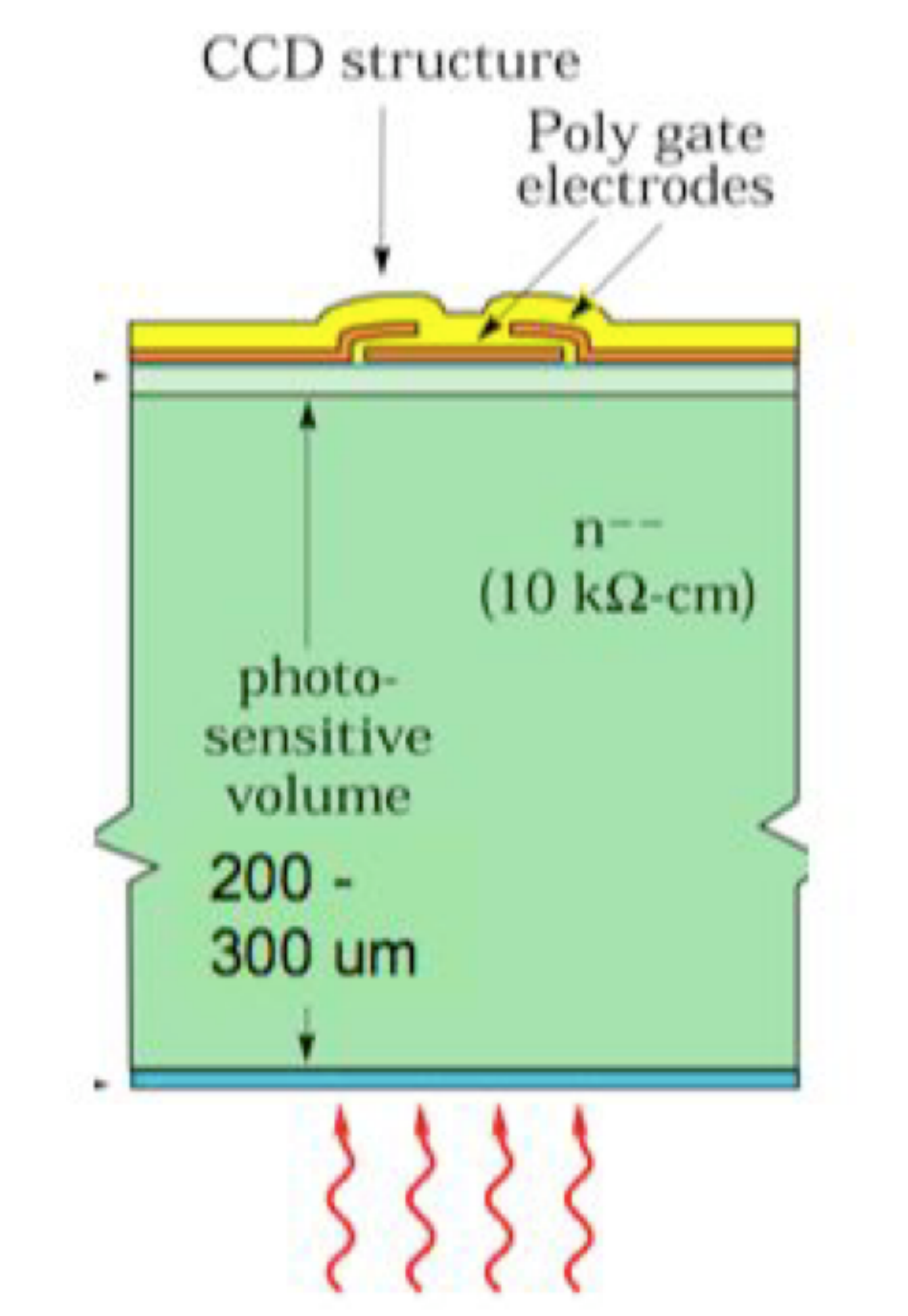}
\end{center}
\caption{Schematic of  a DECam detector. Back illuminated, 250 $\mu$m thick,  p-channel CCD.
For more details see Ref. \cite{LBNL1}}
\label{fig:DESCCD}
\end{figure}

A cartoon of the devices developed  by 
Lawrence Berkeley National Laboratory (LBNL) \cite{LBNL1}  that are 
going to be used in the DECam focal  plane is presented in Fig. \ref{fig:DESCCD}. 
It is a back illuminated,  p-channel CCD thinned to 250 $\mu$m 
and biased from the back side to be fully depleted.
The charge collected in the depletion region is stored in 
the buried channels  established a few
$\mu$m away from the the gate electrodes. 
The holes produced near the back surface must travel the full thickness of the device to reach the 
potential well. During this transit inside the depletion region, 
a hole could also move in the direction perpendicular to the pixel boundaries. 
This effect, called charge diffusion,  has to be kept under control in order to avoid a significant 
degradation in the image quality. The CCDs used in most astronomical instruments until
now are thinned to $< $40 $\mu$m to reduce the charge diffusion. For the DECam CCDs,  
a substrate voltage of up to 80 V is applied to the back surface to 
control diffusion and obtain acceptable image quality in 250 $\mu$m
detectors.

The main features that make the DECam detectors good candidates for a
direct dark matter search are: i) their low electronic noise and ii) their thickness. 
These two features combined would allow building an experiment with significant mass 
(given by the large thickness) and with a very low threshold for nuclear recoils.

\begin{figure}
\begin{center}
\includegraphics[width=0.8\columnwidth]{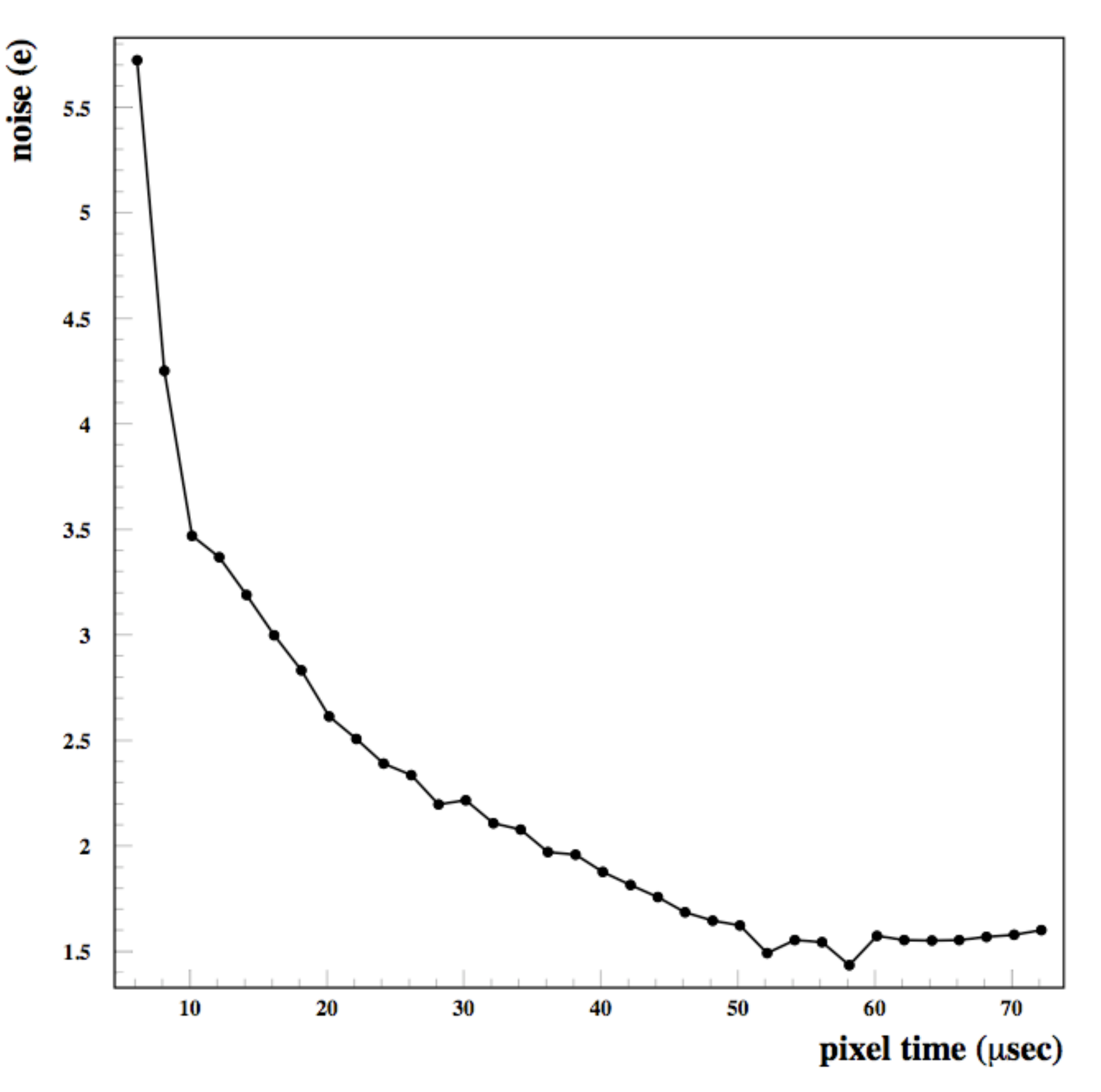}
\end{center}
\caption{Noise as a function of pixel readout time for DECam CCDs. At slow readout speeds
a noise below $\sigma = 2$ e is achieved. }
\label{fig:noise}
\end{figure}

The noise performance of these detectors has been studied in detail as part
of the characterization effort done by the DECam CCD 
team \cite{DECam_CCDs} \cite{DECam_CCDtest}. 
The detectors have 8 million, 15 $\mu$m square pixels and are read in by 
two amplifiers in parallel, each amplifier sitting on opposite ends of a serial register towards
which the charge is clocked. The signal is digitized after correlated double
sampling (CDS) of the output. Each sample used for the CDS operation is
the result of an integration during a time $\tau$. This integration acts
as a filter for high frequency noise. The noise measured for  a DECam detector
as a function of readout time is shown in Fig. \ref{fig:noise}. The noise
observed for pixel readout times larger than 50 $\mu$s is $\sigma < 2e$ (RMS).
At large readout times $\tau$ is one half of the pixel readout time.
The detectors have an output stage with a electronic gain of $\sim 2.5$ $\mu$V/e.
These results were obtained using a Monsoon \cite{monsoon} CCD controller.

\begin{figure}
\begin{center}
\includegraphics[width=0.8\columnwidth]{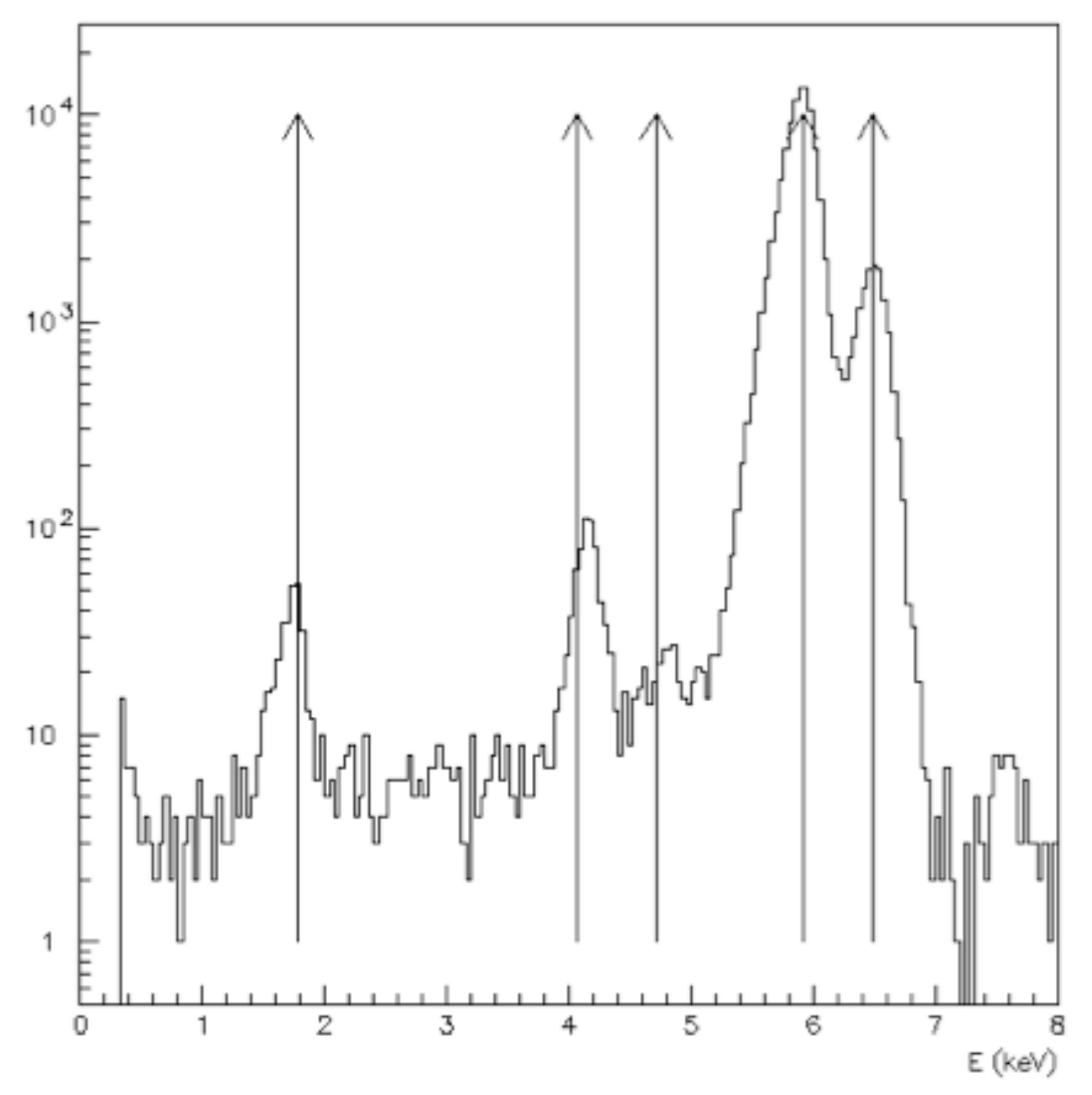}
\end{center}
\caption{Spectrum obtained for the reconstructed X-ray hits in an 
$^{55}$Fe exposure of a DECam CCD.  The arrows mark
the direct X-rays from the source K$\alpha$=5.9 keV and K$\alpha$=6.5 keV,
the   K$\alpha$ and  K$\beta$ escape lines at 4.2 and 4.8 keV, and
the Si X-ray at 1.7 keV.  The  factor
3.64 eV/e is used to convert from charge to ionization energy.}
\label{fig:xrays}
\end{figure}

A commonly used tool in the characterization of CCD detectors is low
energy X-rays from an  $^{55}$Fe source\cite{janesick}. We present here 
the results obtained with $^{55}$Fe in DECam CCDs to demonstrate their performance.  The main
emission of the $^{55}$Fe source is a 5.9 keV X-ray. By reconstructing the 
ionization charge produced by the X-ray  hits,  one can measure
a conversion factor between charge and energy.
This factor can be used to translate the noise measured in units of charge 
to the noise in units of energy for X-ray ionization. This conversion factor is 
known to be 3.64 eV/e- \cite{janesick} and has been 
measured extensively for CCD detectors in general.
An example of the energy spectrum measured for an  $^{55}$Fe 
X-ray exposure in a DECam CCD is shown in Fig. \ref{fig:xrays}, 
which confirms the conversion factor.

\begin{figure}
\begin{center}
\includegraphics[width=0.8\columnwidth]{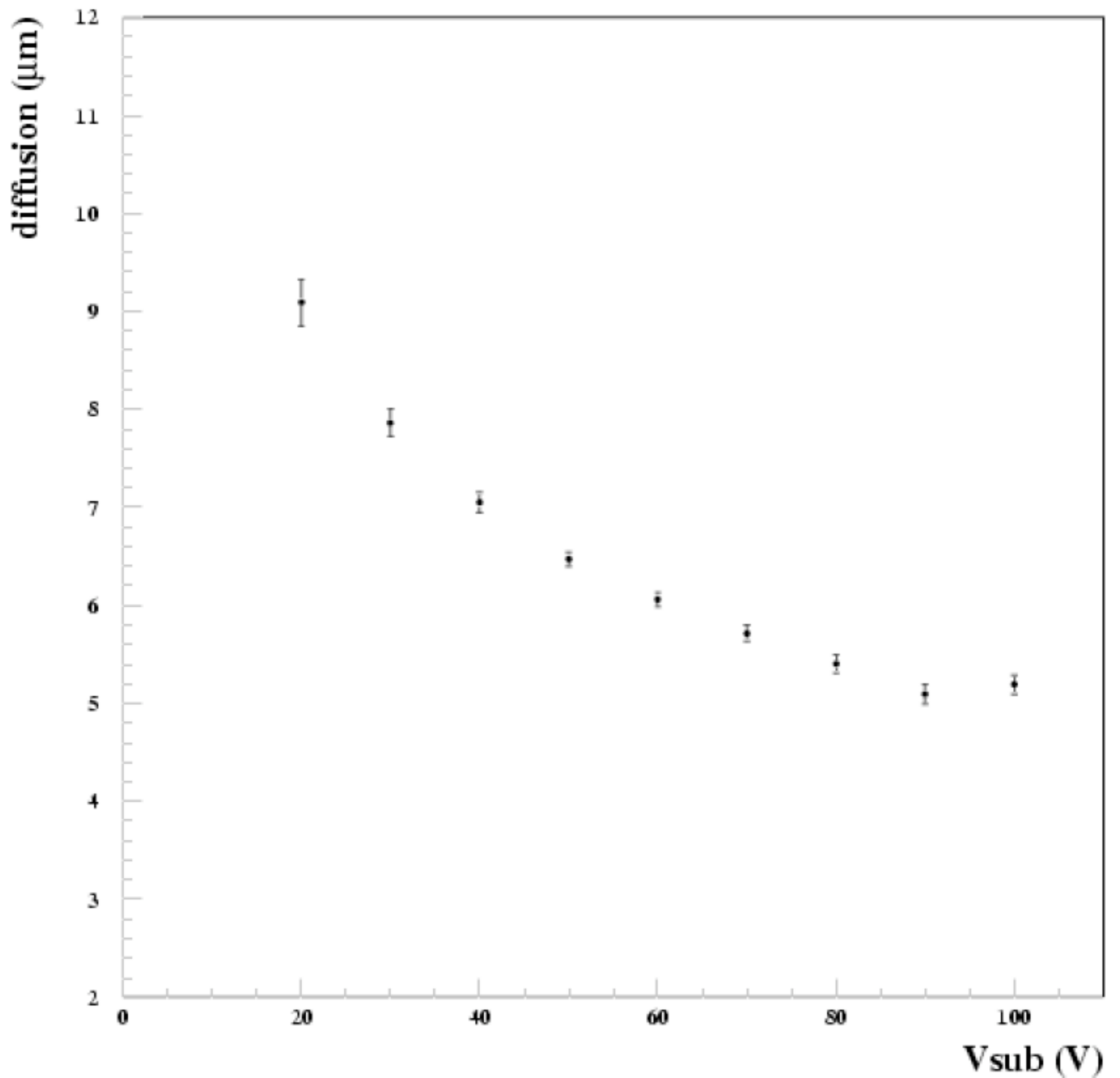}
\end{center}
\caption{Diffusion measurement using X-rays from a $^{55}$Fe source in a  250 $\mu$m DECam CCD with
pixel size 15 $\mu$m x 15 $\mu$m. }
\label{fig:xray_diff}
\end{figure}

The study of  the size of X-ray hits in a back illuminated CCD provides a
measurement of the diffusion. Since the
detectors are back illuminated, a 5.9 keV X-ray produced by an $^{55}$Fe source
will penetrate only about 20 $\mu$m into the silicon before producing
a charge pair \cite{janesick}. The charge produced will have to travel most
of the Si thickness before it can be stored under the potential well for later readout.
As a result of this process, $^{55}$Fe X-rays will produce diffusion limited hits in the detector
corresponding to a known energy deposition.
When considering these detectors for a dark matter search,
diffusion is an important parameter because it  determines
the size of the reconstructed nuclear recoil events.  The nuclear recoils
will produce a very localized charge cloud ( $ \ll 15$ $\mu$m ), and the
signature in the CCD detectors will be a diffusion limited charge deposition,
similar to the X-ray hits.
The diffusion measurement for DECam CCDs 
using X-rays is shown in Fig. \ref{fig:xray_diff}.
This measurement done with X-rays is consistent
with measurements done with optical methods on the same detectors  
\cite{DECAM_diff}, and on other thick detectors developed by LBNL
\cite{LBNL_diff} \cite {LBNL_diff2}.

\section{Nuclear recoil detection with CCDs}

\begin{figure}
\begin{center}
\includegraphics[width=1.0\columnwidth]{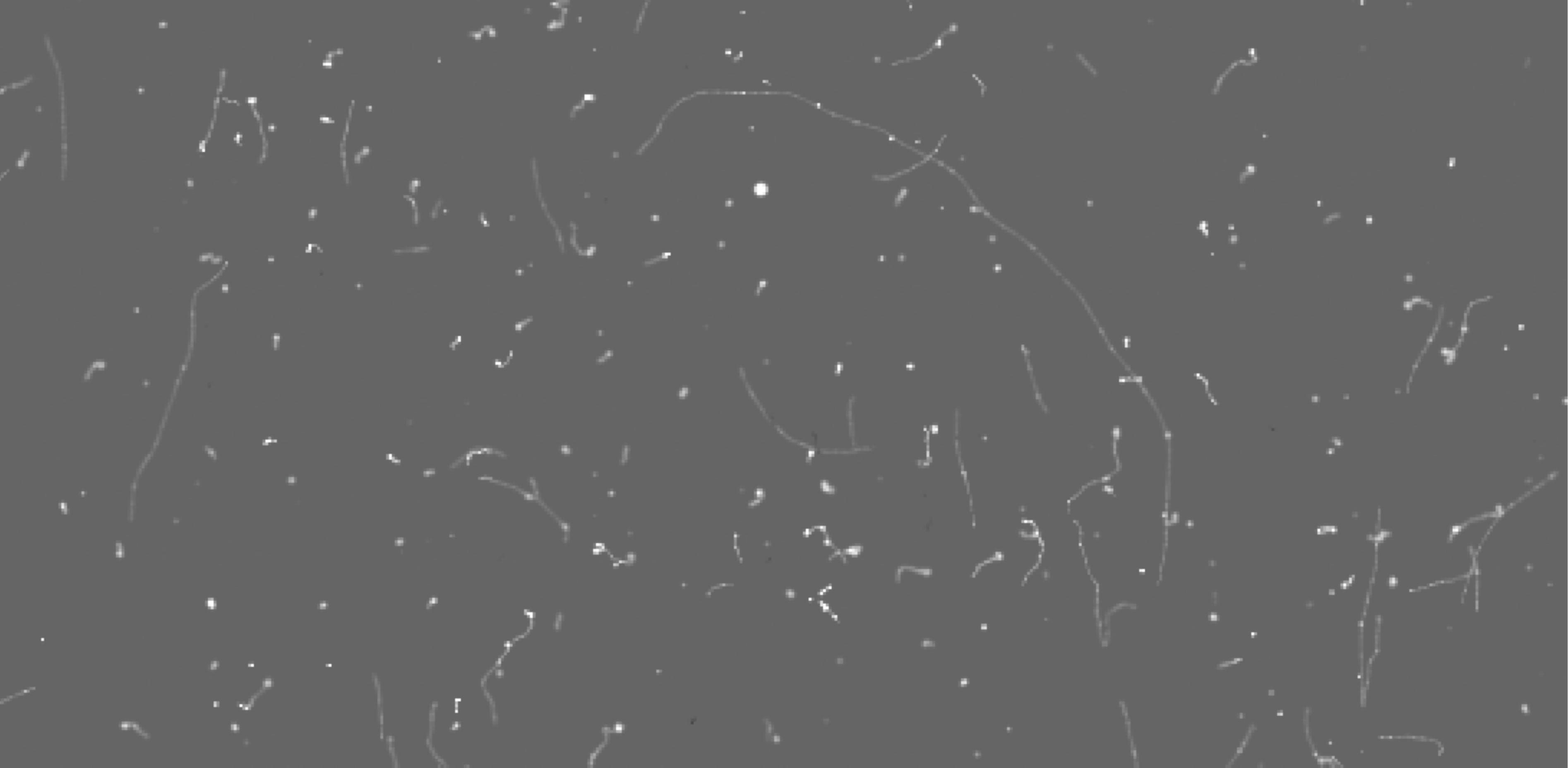}
\end{center}
\caption{Image resulting from an exposure of a DECam CCD to an $^{252}$Cf neutron source.
The total width of the image corresponds to 1000 pixels.}
\label{fig:source}
\end{figure}

In order to demonstrate the nuclear recoil detection with CCDs, a DECam 
detector was exposed to a $^{252}$Cf neutron source, an example of 
the resulting images is shown in Fig. \ref{fig:source}. The image shows
ionization tracks that can be up to a few hundred pixels long 
and diffusion limited hits, typically occupying 
only a few pixels. The nuclear recoil candidates  are the 
diffusion limited hits and are selected according to the diffusion measured 
for these detectors shown in Fig. \ref{fig:xray_diff}. The image was
produced using a substrate voltage of 80 V.  Shape parameters for the
hits are measured using the astronomical image analysis package 
SExtractor \cite{sextractor}, and only those
hits with a principal axis  smaller than 1 pixel are selected as 
nuclear recoil candidates. We used the X-ray data to prove
that this selection is efficient for separating diffusion limited
hits from the rest of the ionization events in the Si.

\begin{figure}
\begin{center}
\includegraphics[width=1.\columnwidth]{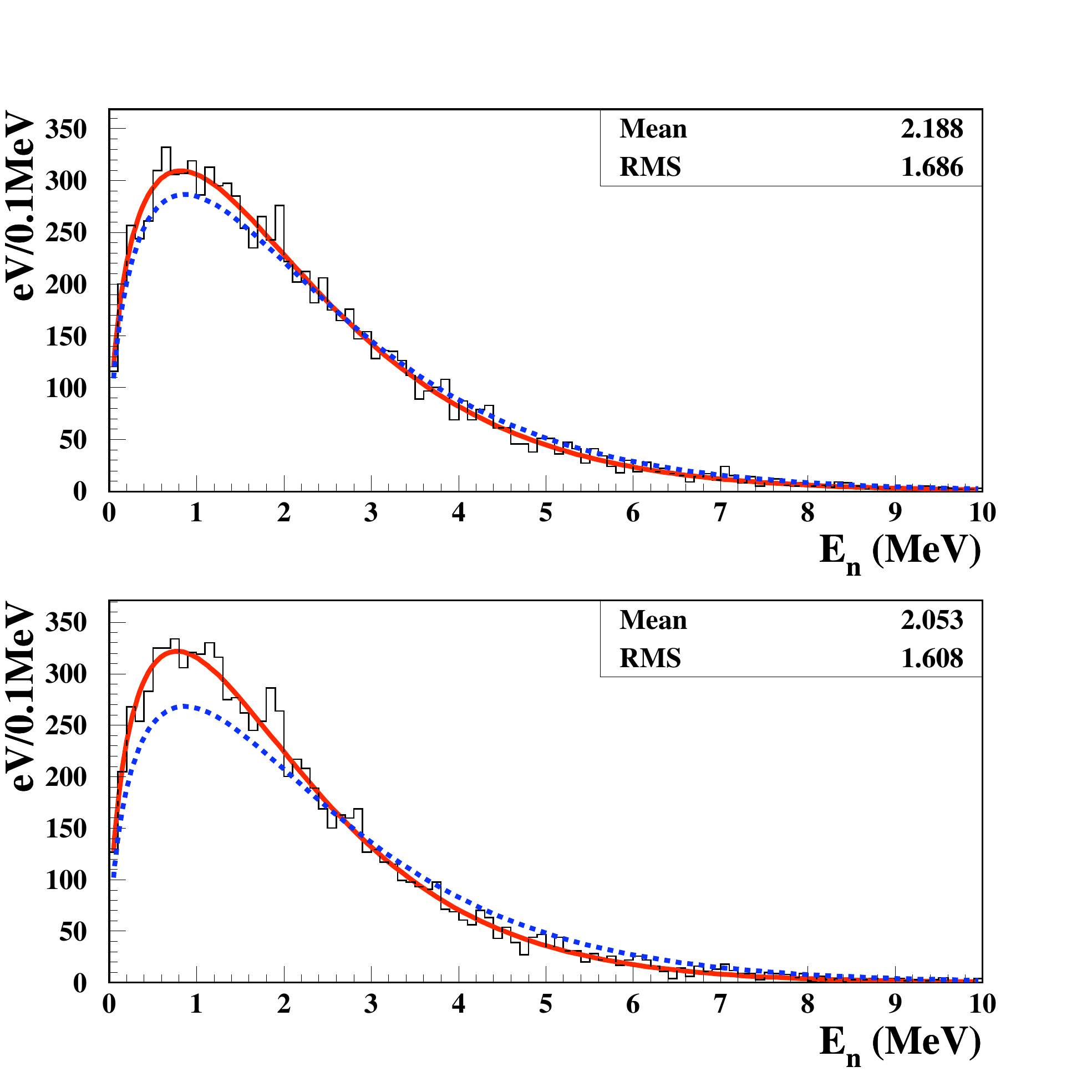}
\end{center}
\caption{Spectrum of neutrons inside the dewar obtained using Geant4 simulation (black histogram). The blue dashed line shows the best fit to Eq.(\ref{eq:cf_spec}) with $\alpha = 0.88$ and $\beta =2$, the red curve is the best fit to Eq.(\ref{eq:cf_spec}) with free $\alpha$ and $\beta$.}
\label{fig:spec_dewar}
\end{figure}

\begin{figure}
\begin{center}
\includegraphics[width=1\columnwidth]{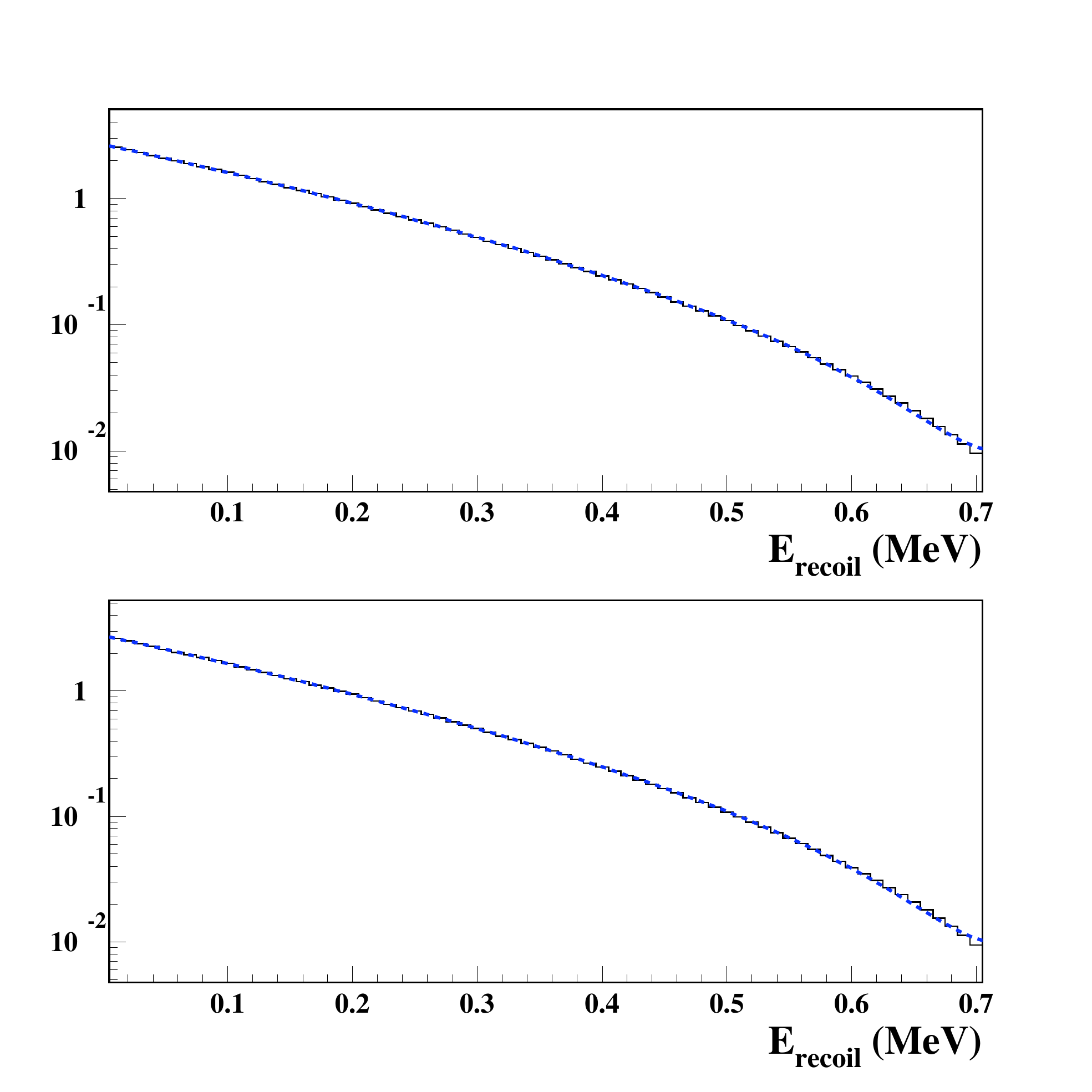}
\end{center}
\caption{Spectrum of nuclear recoils expected according to the simulation for 2.5 cm wall thickness (top) and
5.0 cm wall thickness (bottom).}
\label{fig:recoilsim}
\end{figure}

The energy spectrum from the $^{252}$Cf  neutron source has been measured in previous 
work \cite{CF252spec}, where it was shown to be properly described by the function
\begin{equation}
N (E) = N_0 \exp(-\alpha E) \sinh{\sqrt{ \beta E}}, \label{eq:cf_spec}
\end{equation}
where $N(E)$ is the number of neutrons emitted with kinetic energy $E$,
$N_0$ is a normalization constant and the parameters $\alpha = 0.88$ and
$\beta = 2$.
The DECam detectors used in this work operate at 
-100$^{\rm o}$ C and, for this reason, they are inside
a vacuum dewar with 2.5 cm thick Al walls. To understand
the spectrum of the neutrons inside the dewar, we ran a Geant4 \cite{geant4} simulation.  
The expected spectrum for neutrons inside the dewar as calculated with Geant4 is shown in 
Fig. \ref{fig:spec_dewar}. The simulation was performed for  
a wall thickness  of 2.5 cm and 5.0 cm. The thicker case was considered for neutrons crossing the
Al wall with a large angle.  The expected neutron spectrum inside the dewar can also be described by
Eq.(\ref{eq:cf_spec}) with somewhat different parameters. The spectrum parameters
for the different dewar wall thickness are shown in Table \ref{tab:alpha}.
Once the spectrum of the neutrons inside the dewar is calculated 
as shown in Fig. \ref{fig:spec_dewar}, the energy distribution of the nuclear
recoils produced by a beam of neutrons with energy distribution given by $N(E)$ in
Eq.(\ref{eq:cf_spec}) can be expressed as
\begin{equation}
P(E_{r})  = P_0 \int_{E_{r}}^{\infty} N (E) F(q) {\rm d}E \label{eq:recoil}
\end{equation}
where $E_{r}$ is the recoil energy, $P_0$ is a normalization factor 
and $F(q)$ is the nuclear form factor correction
given as a function of the momentum transfer, see Ref. \cite{formfactor} for details.
The results of the nuclear recoil energy distribution  calculated in Eq.(\ref{eq:recoil} )
are shown in Fig. \ref{fig:recoilsim}, in order to compare them with the spectrum
observed in the CCD the data is fitted to a fourth order polynomial 
\begin{equation}
P(E_r) = A_0+A_1 E_r + A_2 E_r^2 + A_3 E_r^3 +A_4 E_r^4  \label{eq:poly}
\end{equation}
with the parameters shown in Table \ref{tab:poly}.

\begin{table}[h]
       \centering 
        \begin{tabular}{|c|c|c|c|}
         \hline 
           parameter         &   $^{252}$Cf  & 2.5 cm Al & 5.0 cm Al   \\
          \hline  
           \hline
	$\alpha$     & 0.88 	&   0.98 $\pm$ 0.02 & 1.10 $\pm$ 0.03     \\
         $\beta   $     & 2.0	&   2.5    $\pm$  0.3 &  3.1  $\pm$     0.5     \\

                   \hline
                   \end{tabular}
                \caption{Spectrum parameters in Eq.(\ref{eq:cf_spec}) for $^{252}$Cf spectrum, and after the neutrons
                crossed an Al wall of 2.5 cm and 5.0 cm.              }
                \label{tab:alpha}
 \end{table}

\begin{table}[h]
       \centering 
        \begin{tabular}{|c|c|c|c|}
         \hline 
           paremeter         &   2.5 cm Al & 5.0 cm Al   \\
          \hline  
           \hline
         A0   &    2.7    $\pm$      0.2    &       2.8     $\pm$   0.2 \\
         A1   &   -13.0  $\pm$     0.6     &   -13.5     $\pm$   0.6 \\
         A2   &    25.5  $\pm$     1.0     &     26.5     $\pm$  1.0  \\
         A3   &   -24.1  $\pm$     1.6     &   -25.0     $\pm$  1.6 \\
         A4   &    9.2    $\pm$     1.8      &     9.5      $\pm$  1.8 \\
                   \hline
                   \end{tabular}
                \caption{ Parameters for the polynomial fit to the nuclear recoil 
                spectrum in Fig. \ref{fig:recoilsim}.              }
                \label{tab:poly}
 \end{table}

The ionization
efficiency for nuclear recoils is not the same as that for X-rays, as has been
demonstrated in previous work \cite{quench1992}. For this reason
the energy scale for CCDs determined with X-rays will be different
to that for nuclear recoils. For the
the calibration at low recoil energies, relevant to a low threshold experiment, 
ideally one would use  low energy neutrons from a monoenergetic
beam as discussed in Ref. \cite{collarcalib}. This calibration has not
been done for DES CCDs. In this work we will compare
the CCD data obtained with a $^{252}$Cf neutron source, to the expected
recoil spectrum by fitting the ionization efficiency.

\begin{figure}
\begin{center}
\includegraphics[width=1\columnwidth]{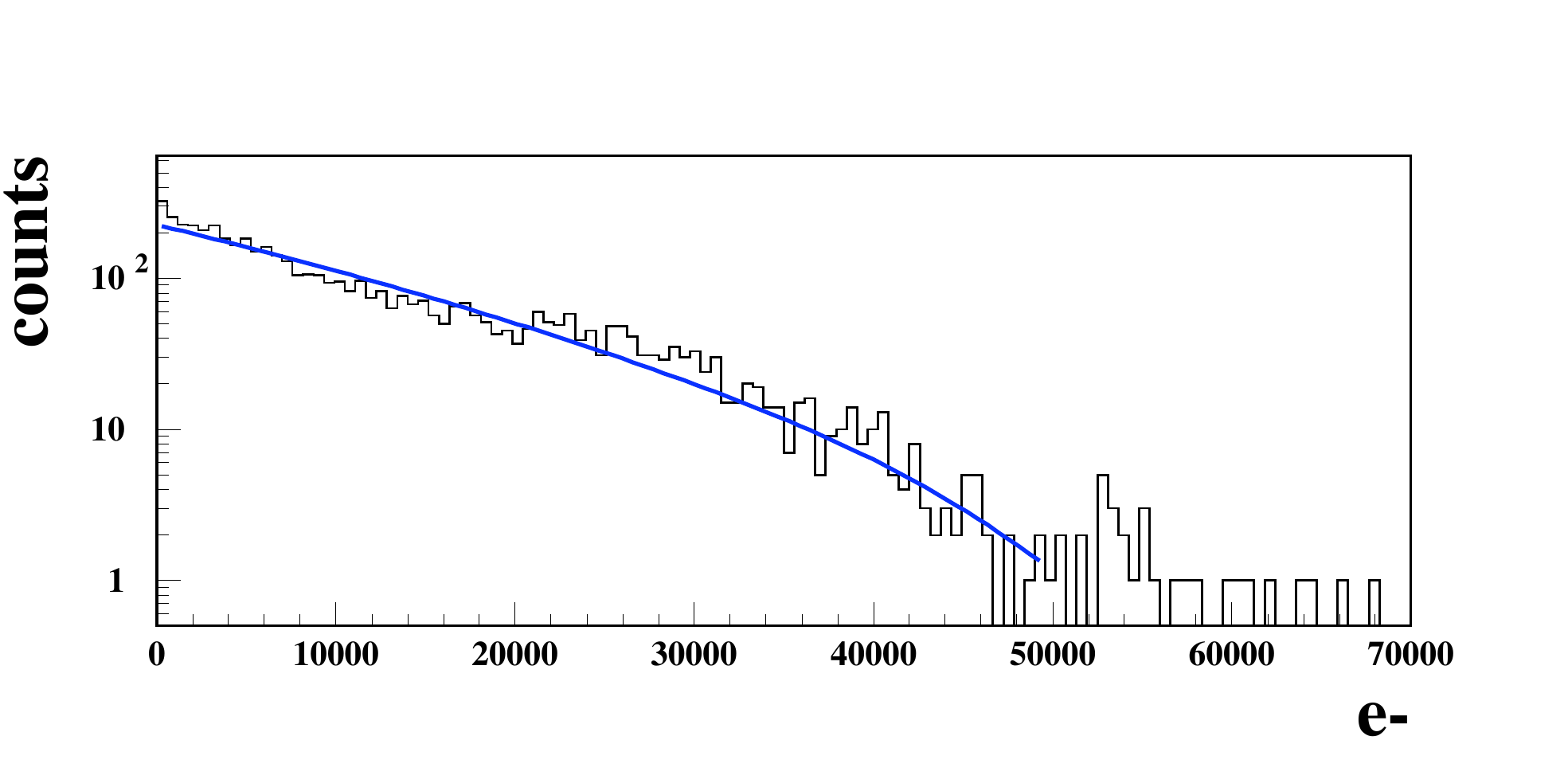}
\end{center}
\caption{Measured spectrum for nuclear recoil candidates in a DECam CCD
exposed to an $^{252}$Cf source. The black histrogram shows the data, and the blue
curve is a fit to $P(fE_r)$ in Eq.(\ref{eq:poly}). Fit to 2.5 cm wall. }
\label{fig:fitdata}
\end{figure}

The expected distribution for the energy of nuclear recoils is compared 
with the data collected for the nuclear recoil candidates in the DECam CCD
exposures, as shown in Fig. \ref{fig:source}.
The comparison is done by fitting  Eq. (\ref{eq:poly}) to the observed spectrum
with a free parameter $f$, to convert the energy of the recoil $E_r$ into measured
charge, effectively fitting $P (f E_r)$ to the data.  The measured spectrum of nuclear
recoil candidates, together with the best fit to $P (f E_r)$ , are shown in Fig. \ref{fig:fitdata}.
The fitted parameter $f$ is shown in Table \ref{tab:fitdata}. Using this fit, we calculate
$Q$ as the ratio between the charge collected for a nuclear recoil and an X-ray
of the same energy, $Q$ is also shown in Table \ref{tab:fitdata}.  '

In previous work \cite{Lindhard} \cite{quench1992} is has been demonstrated that $Q$ depends 
on energy. The data collected in this test is not good enough to study the energy dependency,
however the results obtained for $Q$ are compatible with the measurements done 
in previous work \cite{quench1992}.

\begin{table}[h]
       \centering 
        \begin{tabular}{|c|c|c|c|c|}
         \hline 
           wall thickness  & Vsub        &   f  (e/MeV)& $ \chi^2$/n.d.f. & Q   \\
          \hline  
           \hline
  	  2.5 cm & 80 V &   	74083  $\pm$ 1034	&  2.6    	& 3.71 $\pm$ 0.05	\\
	 5 cm & 80 V     &  	68934  $\pm$ 1047	&  2.7   	& 3.98  $\pm$ 0.06	\\
                        \hline
                   \end{tabular}
                \caption{ Parameters for the polynomial fit to the nuclear recoil 
                spectrum in Fig. \ref{fig:recoilsim}.  The results assuming a 5cm thick wall are also
                presented.            }
                \label{tab:fitdata}
 \end{table}

\section{Plans for CCD Experiment at Low Background (CELB)}

\begin{figure}
\begin{center}
\includegraphics[width=1\columnwidth]{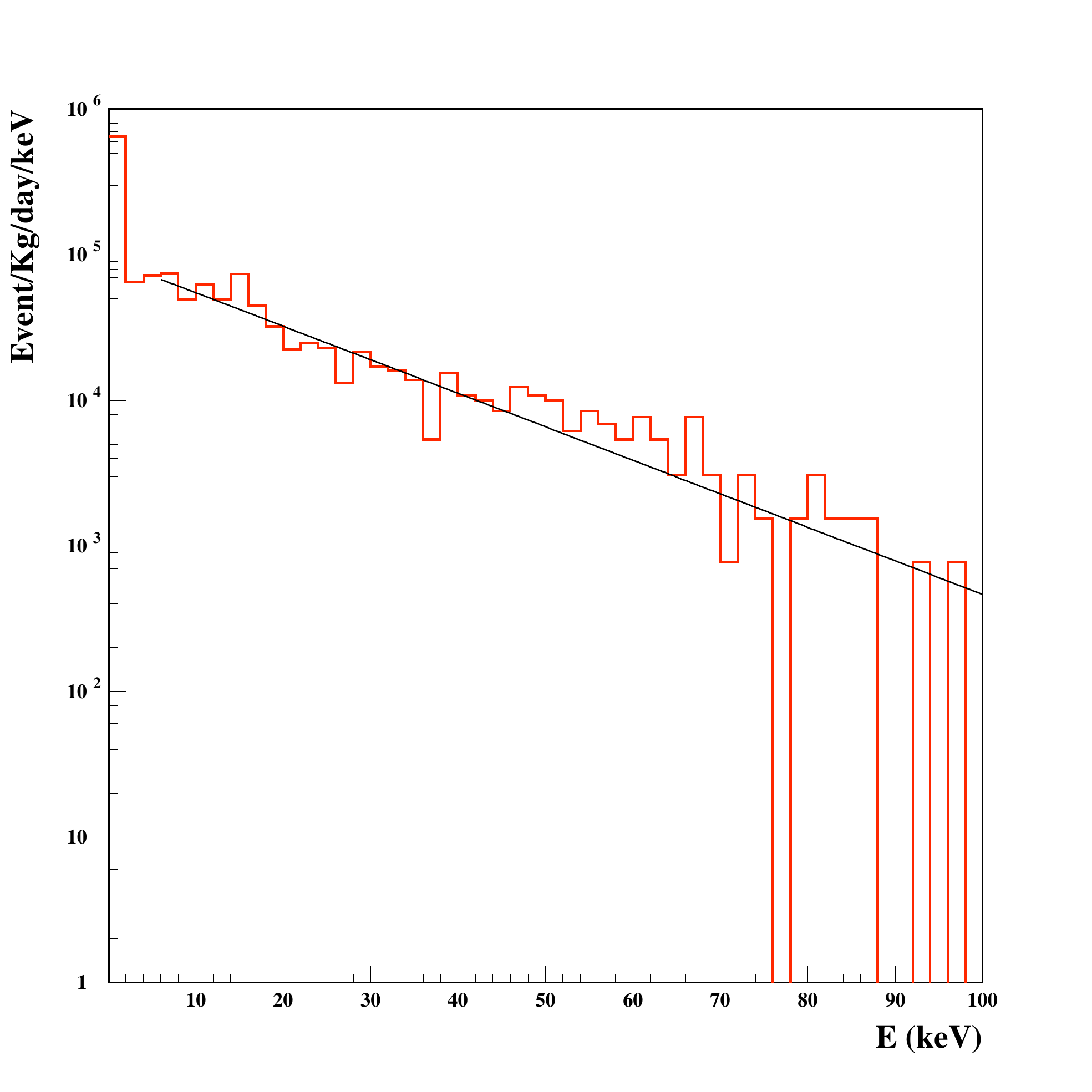}
\end{center}
\caption{Backgound spectrum measured for an unshielded DECam CCD at FNAL (sea level).
The line is an exponential fit $f_1(E) = \exp(a+b E)$, with $a=11.44$ and $b=-0.053$. }
\label{fig:insidelead1}
\end{figure}

\begin{figure}
\begin{center}
\includegraphics[width=1\columnwidth]{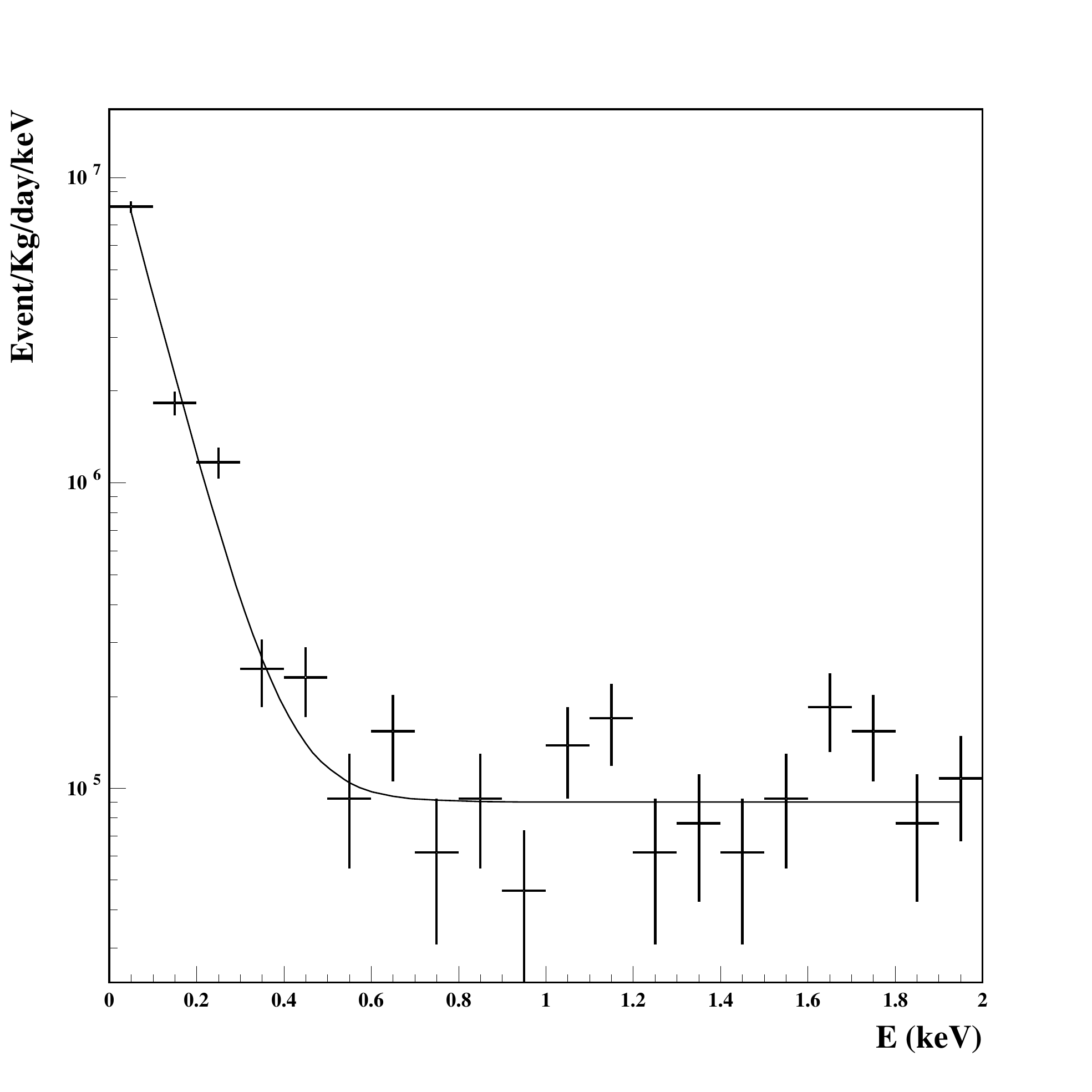}
\end{center}
\caption{Same as Fig. \ref{fig:insidelead1}, but now showing low energy region.
For this region of the energy spectrum an additional 
term is added, the curve corresponds to $f_2(E) = f_1(E) + \exp(c+d E)$,}
\label{fig:insidelead2}
\end{figure}

The detectors discussed above show features that make them ideal
for a CCD Experiment at Low Background (CELB), to conduct a 
direct dark matter search optimized for low mass dark matter candidates.
The $\sigma = 2$e of readout noise measured in the detectors translates
to a $\sigma = $27 eV for
nuclear recoils.   The mass of each DECam CCD is 1 g ( 18 cm$^2$ and 250 $\mu$m thick)
and we envision a 10 g array to be operated underground in the near future.
In the meantime, we collected data for 2~g-day exposure of a
DECam CCD, and the spectrum is shown in Fig. \ref{fig:insidelead1} and 
Fig. \ref{fig:insidelead2}.

A study of radiation events in astronomical images was
presented in Ref. \cite{CCDbackstudy} for thick CCDs, similar
to the DECam detectors. The rate of diffusion limited hits,
characterized as ``spots" in Ref. \cite{CCDbackstudy}, was  measured
to be $\sim$1 cm$^{-2}$ min$^{-1}$ for an unshielded detector at sea level.
These measurements include the full 
dynamic range of the CCDs, corresponding  to energies up to 400 keV.
In these units, we measured 0.8 cm$^{-2}$ min$^{-1}$ for our
unshielded  detectors at Fermilab, which means that our results
are consistent with those presented in   \cite{CCDbackstudy}.

\begin{figure}
\begin{center}
\includegraphics[width=1\columnwidth]{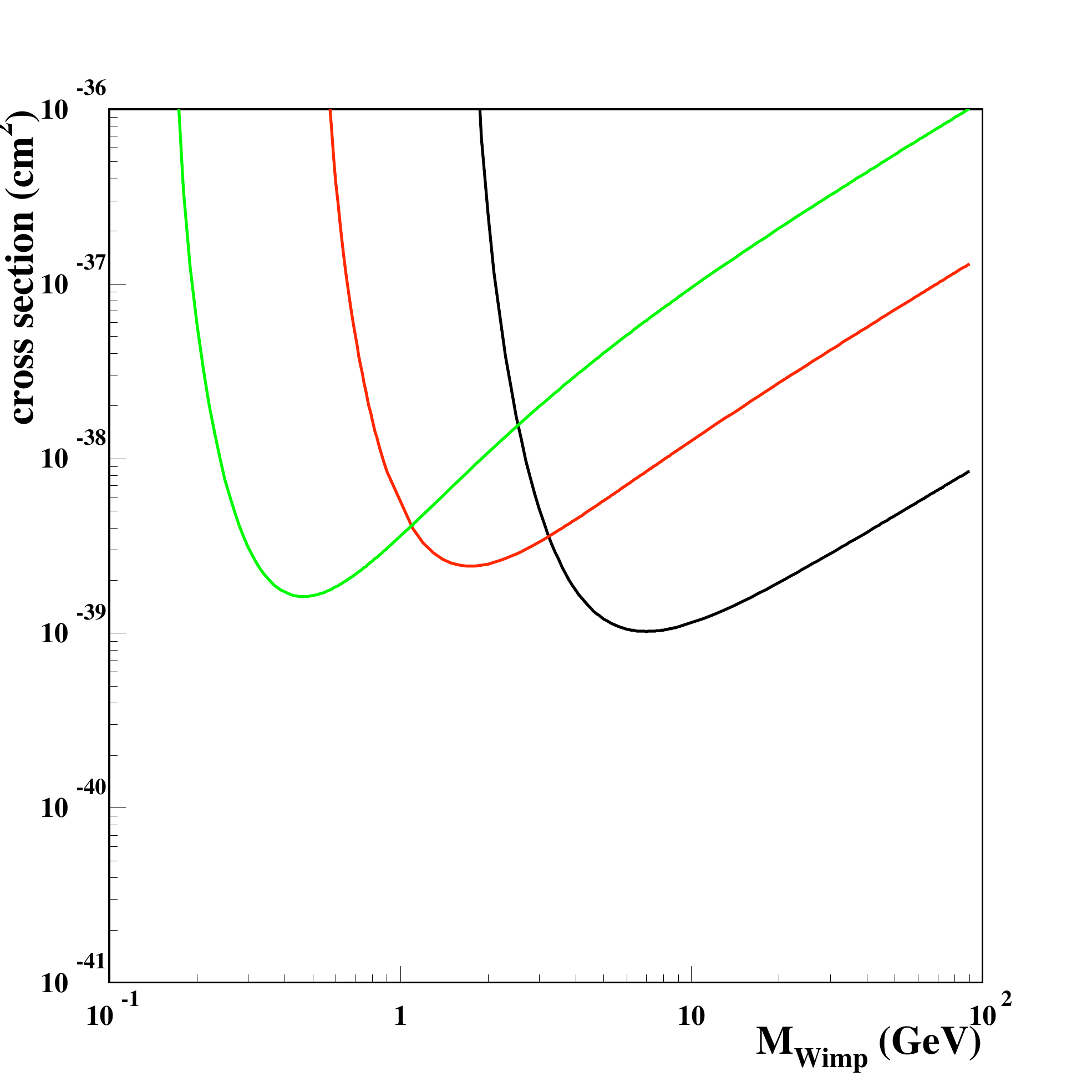}
\end{center}
\caption{Cross section limit expected for a 300 g-day exposure of DECam CCDs underground to be
done during 2008. Red) with a threshold set at  13.5 eV, Green) with threshold set at 135 eV. Black) 
reach with a 1.35 keV threshold is shown for comparison. }
\label{fig:expected}
\end{figure}

The current best limit published for WIMP masses below 5 GeV
corresponds to Ref. \cite{texono} giving a limit for the spin
independent nucleon-WIMP cross section $\sigma < 10^{-39}$ cm$^2$. This limit was established with 
a threshold of $\sim$600 eV  for nuclear recoils
and an exposure of 338 g-day. We expect to
have a 300 g-day exposure underground during 2008 with a
threshold of 135 eV (corresponding to 5 sigma of the electronic noise).  
To estimate the reach of a CCD based search
for dark matter, we assume that the background observed in our 2 g-day run
can be reduced by  4 orders of magnitude by going to an underground facility and building a proper
shield around our detector.  The expected cross section
upper limits for a 300 g-day exposure  are shown in Fig. \ref{fig:expected} .
The results expected with a threshold ten
times higher are also shown for comparison. 

Among the astronomical community the possibility of developing a zero-noise
CCD readout system has been extensively discussed \cite{zeronoise1} \cite{zeronoise2}.
In addition, devices using an electron multiplication stage (CCDEM)  giving 
zero noise performance have been fabricated, although much
thinner than the high resistivity detectors considered here (see for 
example the L3 Vision products from e2v \cite{l3vision}).   A zero readout noise 
detector will allow setting a threshold for 
nuclear recoils at 14 eV (corresponding to 1e), the reach of an 
experiment with such a low threshold is also shown in  Fig. \ref{fig:expected}.

\section{Conclusion}

The prospects for a direct dark matter search using CCDs have been discussed,
demonstrating that the new high resistivity  detectors fully depleted with
a thickness of 250 $\mu$m are good candidates for such an experiment. 

A demonstration of the technology with a 10 g CCD array is planned 
using the DECam CCDs, and we expect that this experiment will set the best limit 
for WIMPs of masses below  5 GeV. The development of a zero noise CCD 
readout could extend this reach below the 1 GeV region.
We believe that an experiment of these characteristics would be a good complement
to the currents dark matter searches because of its extremely low threshold.

\section*{Acknowledgments}

We thank K. Kuk for operating the CCD testing lab where most of these measurements
were performed, the DECam front end electronics group for their
support with this project, and  J. Marriner loaning 
some hardware needed for this work. We thank the LBNL CCD 
team for very useful discussions about the  operation of the DECam detectors
produced in their lab.  Finally, this work would  not have been possible 
without the support  of the Technical Centers  Group in Fermilab's Particle Physics Division. 

\begin{thebibliography}{}

\bibitem{concordance} A. Kogut et al. , The Astrophysical Journal, {\bf 665}    355 (2007).

\bibitem{damalibra} Bernabei, R et al. , arXiv:0804.2741

\bibitem{ddm1} R. J. Gaitskell, Annu. Rev. Nucl. Part. Sci.,  {\bf 54}  315 (2004).

\bibitem{ddm2} G. Chardin, "Cryogenic Particle Detection", edited by Christian Enss, Springer (2005), arXiv:astro-ph/0411503

\bibitem{susy_DM} B. W. Lee and S. Weinberg, Phys. Rev. Lett. , {\bf 39} 165 (1977).

\bibitem{lightDM0} D. Hooper and K.M. Zurek, 	arXiv:0801.3686v1 

\bibitem{lightDM1} X. He, T. Li, X. Li and H. Tsai,  Modern Physics Letters A, {\bf 22}  2121 (2007).

\bibitem{lightDM5} J.F. Gunion, D. Hooper, B.  McElrath, Phys.Rev., {\bf D73} 015011 (2006) .


\bibitem{lightDM4} C. Bird, R. Kowalewski \& M. Pospelov, Modern Physics Letters A,  {\bf 21}  457 (2006).

\bibitem{lightDM2} Gondolo, P. and Gelmini, G.,  Phys. Rev. , {\bf D 71} 123520 (2005).



\bibitem{lightDM3} T. Damour and L.M. Krauss,  
``Proceedings of the 3rd International Workshop on the Identification of Dark Matter", edited by N. J. C. Spooner \& V. Kudryavtsev. World Scientific (2001), arXiv:astro-ph/9806165v3


\bibitem{juancollar}  P.S. Barbeau, J.I .Collar and O. Tench, Journal of Cosmology and Astroparticle Physics, {\bf 09} 009 (2007).



\bibitem{LBNL1}  S.E. Holland, D.E. Groom, N.P. Palaio, R. J. Stover, and M. Wei,  IEEE Trans. Electron Dev.,  {\bf 50}  225  (2003),  LBNL-49992. 


\bibitem{DECam} Flaugher, B., Ground-based and Airborne Instrumentation for Astronomy. Edited by McLean, Ian S.; Iye, Masanori. Proceedings of the SPIE, Volume 6269, (2006)

\bibitem{DES}  Dark Energy Survey Collaboration, astro-ph/0510346.

\bibitem{SNAP}  ``Supernova / Acceleration Probe: A Satellite Experiment to Study the 
  Nature of the Dark Energy'',  SNAP Collaboration, G.~Aldering {\it et al.}, 
  submitted to Publ. Astr. Soc. Pac.,  astro-ph/0405232; SNAP Collaboration, astro-ph/0507459.


\bibitem{Hypersuprime} M. Satoshi et al., Ground-based and Airborne Instrumentation for Astronomy. Edited by McLean, Ian S.; Iye, Masanori. Proceedings of the SPIE, Volume 6269, (2006)


\bibitem{DECam_CCDs}  J. Estrada \& R. Schmidt , Scientific Detectors for Astronomy 2005, Edited by J.E. Beletic, J.W. Beletic and P. Amico, Springer, (2006).

\bibitem{DECam_CCDtest}  J. Estrada et al. , Ground-based and Airborne Instrumentation for Astronomy. Edited by McLean, Ian S.; Iye, Masanori. Proceedings of the SPIE, Volume 6269,  (2006).

\bibitem{janesick} J.R. Janesick, Scientific Charge Cupled Devices, SPIE press (2001).

\bibitem{BlancoCTIO} T. M. C. Abbott et al. ,Ground-based and Airborne Instrumentation for Astronomy. Edited by McLean, Ian S.; Iye, Masanori. Proceedings of the SPIE, Volume 6269, (2006)

\bibitem{DECAM_diff} H. Cease, H. T. Diehl, J. Estrada, B. Flaugher and V. Scarpine,  Experimental Astronomy , Online First (2007)

\bibitem{monsoon}  http://www.noao.edu/ets/new$\_$monsoon 

\bibitem{LBNL_diff}  J.A. Fairfield , D. E. Groom, S. J. Bailey, C. J. Bebek, S. E. Holland, A. Karcher, W. F. Kolbe, W. Lorenzon, \& N. A. Roe, Fairfield IEEE Trans. Nucl. Sci.  53  (6), 3877 (2006) 

\bibitem{LBNL_diff2} A. Karcher, C.J. Bebek, W. F. Kolbe, D. Maurath, V. Prasad, M. Uslenghi, M. Wagner, IEEE Trans. Nucl. Sci. 51 (5), (2004) LBNL-55685.

\bibitem{Lindhard} J. Lindhard, V. Nielsen, M. Scharff, and P.V. Thomsen, Mat. Fys. Medd. Dan. Selsk {\bf 33}, 10 (1963).


\bibitem{quench1992}  B.L. Dougherty, Physical Review A {\bf 45} 2104 (1992)

\bibitem{collarcalib} Barbeau, P. S.; Collar, J. I.; Whaley, P. M., Nuclear Instruments and Methods in Physics Research Section A,  {\bf 574} 385 (2007).

\bibitem{sextractor} Bertin, E. and Arnouts, S. Astronomy and Astrophysics Supplement  {\bf 117}  393 (1996).

\bibitem{diff_Xrays} S.A. Rodney \& J.L. Tonry, astro-ph/0604322.

\bibitem{formfactor}  J.D. Lewin and P.F. Smith, Astropart. Phys. 6, 87 (1996).

\bibitem{diff_textbook}  F.S.Crawford, McGraw-Hill, 1968.


\bibitem{CF252spec}  A.B. Smith and P.R. Fields , Physical Review, {\bf 108} (1957).


\bibitem{geant4}   http://geant4.web.cern.ch/geant4/ 

\bibitem{CCDbackstudy} A. R. Smith, R. J. McDonald, D. L. Hurley, S. E. Holland, and D. E. Groom, SPIEÕs Electronic Imaging (2002), LBNL-49316 


\bibitem{texono}   S. T. Lin et al. (TEXONO Collaboration) arXiv:0712.1645v1 [hep-ex]

\bibitem{zeronoise1} Jean-Luc Gach et al, "Scientific Detectors for Astronomy 2003", Edited by J.E. Beletic, J.W. Beletic and P. Amico, Springer, (2004).
The Beginning of a New Era",  

\bibitem{zeronoise2} Jean-Luc Gach et al,  Publications of the Astronomical Society of the Pacific {\bf115} 1068 (2003).

\bibitem{l3vision}   http://www.e2v.com/ 


\end{thebibliography}



\end{document}